\documentclass[aps,prd,reprint,superscriptaddress,longbibliography]{revtex4-1}
\usepackage{hyperref}
\usepackage{graphicx}
\usepackage{amsmath,amssymb}
\usepackage{subfigure}
\usepackage{color,xcolor}
\usepackage{times}

\begin{document}
 
\title{Scattering of massless scalar waves from Schwarzschild-Tangherlini black holes on the brane}

\author{C\'assio I. S. Marinho}
\email{cismarinho@ufpa.br}
\affiliation{Faculdade de F\'isica, Universidade Federal do Par\'a,
	66075-110, Bel\'em, Par\'a, Brazil}

\author{Ednilton S. de Oliveira}
\email{ednilton@pq.cnpq.br}
\affiliation{Faculdade de F\'isica, Universidade Federal do Par\'a,
	66075-110, Bel\'em, Par\'a, Brazil}

\date{\today}

\begin{abstract}
 The scalar scattering cross section is calculated for Schwarzschild-Tangherlini black holes on the brane. The cross sections are computed numerically via the partial-wave method. The phase shifts are found numerically and also via the Born approximation. It is shown that spacetimes with more than six dimensions present a finite scattering cross section in all directions. 
\end{abstract}


\maketitle

\section{Introduction}

Black holes in General Relativity (and its modified variants) have been a source of fascination ever since the first appearance of an exact solution of Einstein's equations, over a century ago~\cite{Chandra_1983}. In standard four-dimensional spacetimes, black holes are expected to exhibit various phenomena, such as Hawking radiation~\cite{Hawking_1974:nature248_30}, superradiance~\cite{Starobinskii_1973:jetp37_28}, and the interconversion of electromagnetic and gravitational radiation~\cite{Gerlach_1974:prl31_1023}. In some cases, black holes may be endowed with supersymmetric properties~\cite{Oliveira_2011-prd84_084048,Crispino_2015:arXiv1507.03993}. Extreme compact objects, such as black holes, will leave a characteristic imprint on scattered radiation due the presence of the photon sphere~\cite{virbhadra2000prd62_084003,claudel2011jmp42_818} and the event horizon, artifacts of the strong gravitational interaction. This fact has motivated several studies of their scattering properties~\cite{Futterman_etal-1988,Crispino_2009-prl102_231103}.

In recent years, there has been increased interest in black holes in extra-dimensional spacetimes. 
 This is due, in part, to the Arkani-Hamed--Dimopoulos--Dvali (ADD) conjecture, which proposes the existence of large extra dimensions in order to resolve the hierarchy problem~\cite{Arkani-Hamed_1998:plb429_263,AntoniadisPLB436}. This is achieved by taking the electroweak energy scale ($\sim 1$ TeV) to be the only fundamental short distance scale in nature. As the Standard-Model particles and effects are known to be very accurately measured in this scale, ADD had conjectured that only the gravitational field (and other unknown fields) can propagate in the full spacetime -- the bulk -- while the Standard-Model particles are confined in the 3-brane, the usual four-dimensional spacetime.
In scenarios like this, the production of black holes becomes possible in TeV-scale colliders, such as the LHC~\cite{emparan2008lrr2008_6}.

In the present paper we use the ADD conjecture to compute the scattering properties of small Schwarzschild black holes for the massless scalar field propagating on a 3-brane.
The $(n+3)$-dimensional Schwarzschild black holes are described by the metric found by Tangherlini \cite{tangherlini}. This metric describes a spherically symmetric spacetime in $(3+n)$ dimensions. The line element is given by~\footnote{Here we adopt the natural system of units $(G=\hbar=c=1)$ (unless in specific cases where we can explicit them) and use the metric signature $(+,-,\ldots,-)$.}
\begin{equation}
ds^2=g_{AB}^{\mathrm{(bulk)}}dX^A dX^B=fdt^2-f^{-1}dr^2-r^2d\Omega_{n+1},\label{metric-d-s}
\end{equation}
where $X^A$ are the bulk coordinates $\{X^A\}=(t,r,\theta_1,\theta_2,\ldots)$, $d\Omega_{n+1}$ is the $3+n$ dimensional solid angular element. The previous $f$ function is
\begin{equation}
f\equiv f(r)=1-\frac{\mu}{r^{n}},\label{fder}
\end{equation}
where
\begin{equation}
\mu=r_h^n=\dfrac{16\pi G_nM/c^2}{(n+1)\Omega_{n+1}}
\end{equation}
is the mass parameter, $r_h$ is the event horizon radius, $M$ is the black hole mass, $G_n$ is the $(3+n)$ dimensional Newton's constant,  $c$ is the speed of light and $\Omega_N$ is the volume of a unit $N$-sphere.

Here we use an induced Schwarzschild-Tangherlini metric on the brane. Following Ref.~\cite{emparan_2000-prl85_499}, we consider a test 3-brane with negligible self-gravity, where the metric is given by the relation:
\begin{equation}
g_{\mu\nu}^{\mathrm{(brane)}}=g_{AB}^{\mathrm{(bulk)}}\frac{\partial X^A}{\partial x^\mu}\frac{\partial X^B}{\partial x^\nu}.
\end{equation}
 By choosing the brane coordinates as $x^\mu=\delta^{\mu}_{A}X^A$, its induced metric is:
\begin{equation}
	g_{\mu\nu}^{\mathrm{(brane)}}=\textrm{diag}\left(f,-1/f,-r^2,-r^2\sin^2\theta\right).\label{metric-brana}
\end{equation}
This is a spherically symmetric metric \cite{gravitation}, but it is noteworthy that it is not a solution of the four-dimensional Einstein's equations in vacuum.

This paper is organized as follows: in Sec.~\ref{Sec:Classical&Semiclassical} we present some useful analytical results, such as the classical limit and the glory approximation. In Sec.~\ref{Sec:PartialWave} we apply the partial-wave method in the description of the massless scalar field and its general scattering properties and we obtain the phase shifts via the Born approximation. The numerical procedure to obtain the phase shifts is described in Sec.~\ref{Sec:Num.Analysis}. In Sec.~\ref{Sec:Results} we present our results for the cross sections. In Sec. \ref{Sec:Conclusion} we conclude with some final remarks.

\section{Classical and Semi-classical Approximations}\label{Sec:Classical&Semiclassical}

The geodesics which particles on the brane follow can be obtained by the usual geodesic equation from General Relativity using the brane-induced metric \eqref{metric-brana}. Since these spacetimes are invariant under both time translation and rotation around the origin, we have two Killing vectors $\xi=\xi^\mu\partial_\mu=\partial_t$ and $\chi=\chi^\mu\partial_\mu=\partial_\phi$, and therefore two conserved quantities, namely the energy
\begin{equation}
E=g_{\mu\nu}^{(\mathrm{brane})}\xi^\mu\dot{x}^\nu=f\dot{t}
\label{E}
\end{equation}
and the angular momentum
\begin{equation}
 L=-g_{\mu\nu}^{(\mathrm{brane})}\chi^\mu\dot{x}^\nu=r^2\sin^2\theta \dot{\phi}.
 \label{L}
\end{equation}

The spherical symmetry allow us to set $\theta=\pi/2$ without loss of generality. For massless particles, the tangent vector $\dot{x}^{\mu}$ to the geodesic is a null vector, so that $g_{\mu\nu}^{(\mathrm{brane})}\dot{x}^{\mu}\dot{x}^{\nu}=0$. The constants $E$ and $L$ together with the geodesic equation give us the energy-like conservation equation
\begin{equation}
\frac{1}{2}\dot{r}^2+V_{\mathrm{eff}}(r)=\frac{E^2}{2}, \label{energia.sm}
\end{equation}
where
\begin{equation}
V_{\mathrm{eff}}(r)=\frac{L^2f}{2r^{2}}
\end{equation}
is the effective potential. The critical points of this function correspond to unstable orbits with radius $r_c$. One can verify that these orbits are located at
\begin{equation}
r_c=r_h\left(\frac{n+2}{2}\right)^{1/n}.
\end{equation}
The critical impact parameter $b_c$ is obtained by making $E=E_c=\sqrt{2V_{\mathrm{eff}}(r_c)}$, what leads to
\begin{equation}
b_c=\frac{L}{E_c}=r_c\sqrt{\frac{n+2}{n}}.
\end{equation}

Using Eqs.~\eqref{L} and \eqref{energia.sm}, we can write the orbital equation as
\begin{equation}
\left(\frac{du}{d\phi}\right)^2=\frac{r_h^2}{b^2}-u^2(1-u^n),\label{orbital.sm}
\end{equation}
where $u \equiv r_h/r$ and $b=L/E$ is the impact parameter. The direct integration of this equation gives the deflection angle $\Theta(b)=\Delta\phi-\pi$ as function of the impact parameter. In the weak-field limit (in general, low scattering angles), it has been shown via geodesic analysis that \cite{PhysRevD.79.064014}:
\begin{equation}
\Theta(b) \approx \beta_n(r_h/b)^n,
\label{wf_def}
\end{equation}
with $\{\beta_n\}=\{2,3\pi/4,8/3,15\pi/16,16/5,35\pi/32\}$, for $n=1,2,\ldots 6$, respectively. By evaluating the classical scattering cross section via 
\begin{equation}
\frac{d\sigma}{d\Omega}\Big|_{\mathrm{clas}}=\sum_{k}\frac{b(\theta)}{\sin\theta}\left|\frac{db}{d\theta}\right|,\label{eq.class.scat.}
\end{equation}
where the summation in $k$ accounts the number of times the massless particle rotates around the black hole, we can obtain that, in the weak-field limit,
\begin{equation}
\left.\frac{d\sigma_{\mathrm{el}}^{{(n)}}}{d\Omega}\right|_{\theta\approx 0} \approx \frac{r^2_h}{n\theta^2}\left(\frac{\beta_n}{\theta}\right)^{2/n}.
\label{scatter-high-frenq}
\end{equation}

It is interesting to note that this equation implies there is an infinite flux of classical particles scattered in the forward direction for any number of dimensions. However, as we will see below, this only holds in the classical limit. By using phase shifts obtained via the Born approximation, we show that the scattering cross sections in the forward direction for $n \ge 4$ are finite for limited-frequency waves.

In the strong-field regime, an important result for the scattering cross section of black holes predicts the appearance of a `glory' -- concentric rings -- in the backward direction $\theta(b_g)=\pi$. The glory arises as a consequence of particles being scattered near the unstable circular orbit, which occurs when particles impinge upon the black hole with a critical impact parameter, $b_c$. The authors of Ref. \cite{PhysRevD-Matzner} have found an analytical expression to compute the scattering cross section for massless waves near the backward direction:
\begin{equation}
\frac{d\sigma}{d\Omega}\Big|_{\theta\approx\pi}\approx 2\pi\omega b_g^2\left|\frac{db}{d\theta}\right|_{\theta=\pi}[J_{2s}(b_g\omega\sin\theta)]^2,
\label{eq.glory.scat.}
\end{equation}
with $s$ and $\omega$ being the spin and the frequency of the wave, respectively, and $J_\mu(\cdot)$ being the Bessel functions of first kind of order $\mu$ \cite{Gradshteyn}.

Using a numerical approach, we can obtain both $b_g$ and $db/d\theta|_{\theta=\pi}$; the results for the first values of $n$ are listed in Table~\ref{tb:gloryparameters}. From this table we infer that, as $n$ increases, the intensity of the glory rings, given by $2\pi\omega b_g^2\left|db/d\theta\right|_{\theta=\pi}$, decreases and they become wider as $b_g$ decreases.

\begin{table}[htb!]
	\centering
\caption{Parameters of the glory approximation obtained numerically for different dimensions.}\label{tb:gloryparameters}
\begin{tabular}{ccc}\hline\hline
	$n$ & $b_g/r_h$ & $|db/d\theta|_{\theta=\pi}\times\pi/r_h$ \\ \hline 
	1 & 2.6784793 & $0.26790787$ \\ 
	2 & 2.0089973 & $0.40558087\times10^{-1}$ \\ 
	3 & 1.7541535 & $0.10964304\times10^{-1}$ \\ 
	4 & 1.6124635 & $0.38322164\times10^{-2}$ \\ 
	5 & 1.5203409 & $0.15586516\times10^{-2}$ \\ 
	6 & 1.4549229 & $0.70074496\times10^{-3}$ \\ \hline \hline
\end{tabular} 	
\end{table}

\section{Partial Wave Method}\label{Sec:PartialWave}

The Klein-Fock-Gordon equation for the minimally coupled massless scalar field propagating on the 3-brane reads
\begin{equation}
\frac{1}{\sqrt{|g_{(\mathrm{brane})}|}}\partial_\mu\left[\sqrt{|g_{(\mathrm{brane})}|} g^{\mu\nu}_{\mathrm{(brane)}}\partial_\nu\Phi(x)\right]=0.
\end{equation}
Here we consider monochromatic waves. Therefore, we use the following \textit{ansatz} for separation of variables 
\begin{equation}
\Phi(x^\mu)=\frac{1}{r}\psi_{\omega l}(r)Y^{m}_{l}(\theta,\phi)\textrm{e}^{-i\omega t},\label{eq:decomposition}
\end{equation}
where $Y^{m}_{l}(\theta,\phi)$ are the spherical harmonics \cite{abramowitz} and  $\omega$ is the frequency of the wave. The equation for the radial part can be written as
\begin{equation}
f\frac{d}{dr}\left[f\frac{d}{dr}\psi^{(n)}_{\omega l}(r)\right]+\left[\omega^2-V_{l}(r)\right]\psi^{(n)}_{\omega l}(r)=0,\label{r.radial-eq}
\end{equation}
with the effective potential $V_{l}(r)$ given by
\begin{equation}
V_{l}(r)=\frac{f}{r^2}\left[l(l+1)+n\frac{r_h^n}{r^{n}}\right].
\end{equation}
In order to obtain the asymptotic solutions, we adopt a Wheeler-type coordinate, as 
\begin{equation}
r_*=\int^{r}\frac{d\xi}{f(\xi)}+C_n,
\end{equation}
where the constant $C_n$ is chosen such that $r_*\rightarrow r$ when $r\rightarrow\infty$. With this coordinate, Eq.~\eqref{r.radial-eq} is written as
\begin{equation}
\frac{d^2}{dr_{*}^{2}}\psi^{(n)}_{\omega l}(r_*)+\left[\omega^2-V_{l}(r_*)\right]\psi^{(n)}_{\omega l}(r_*)=0,\label{asymtoptic}
\end{equation}
which is similar to the time-independent Schr\"{o}dinger equation. 

\begin{figure}
	\centering
	\subfigure[\ $l=0$.]{\label{pot.r.l0}\includegraphics[width=\columnwidth]{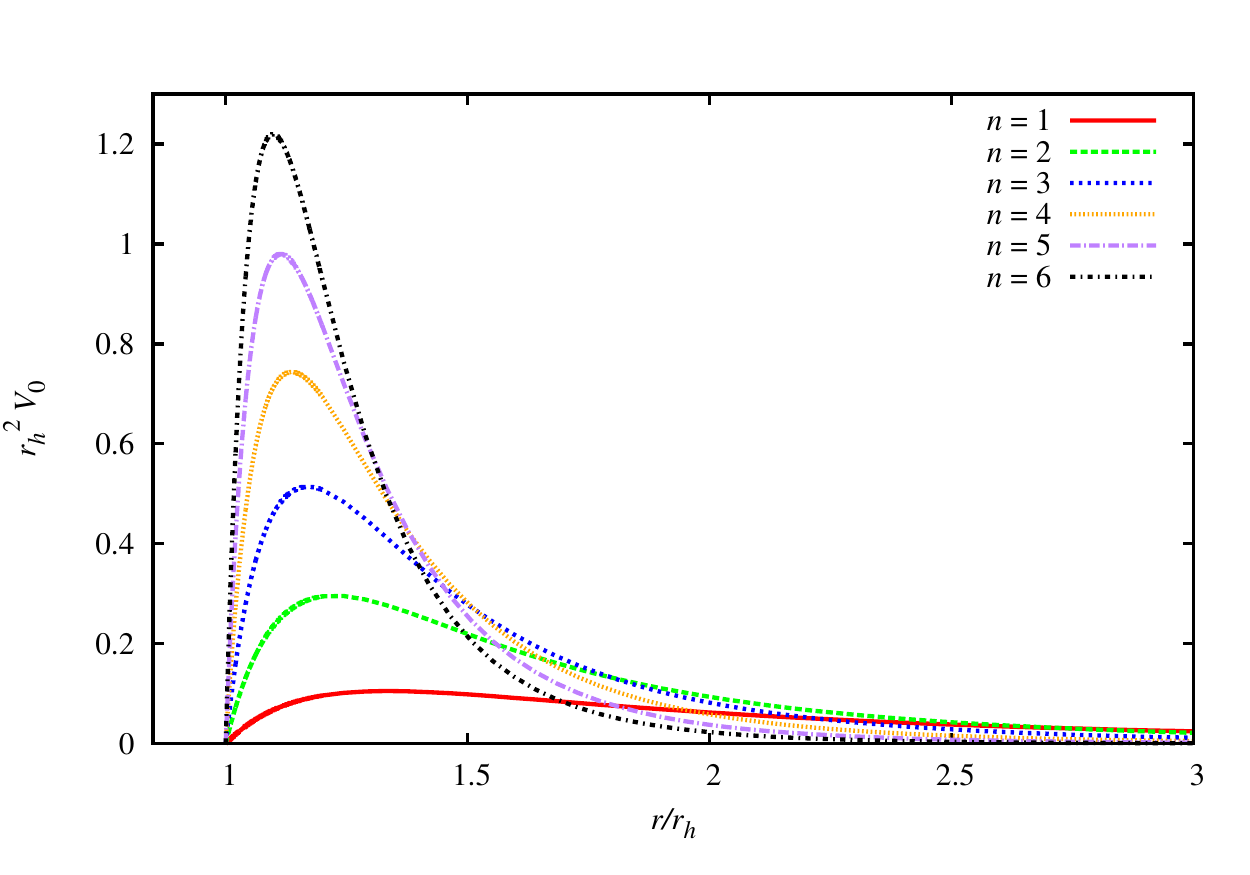}}
	\subfigure[\ $l=2$.]{\label{pot.r.l2}\includegraphics[width=\columnwidth]{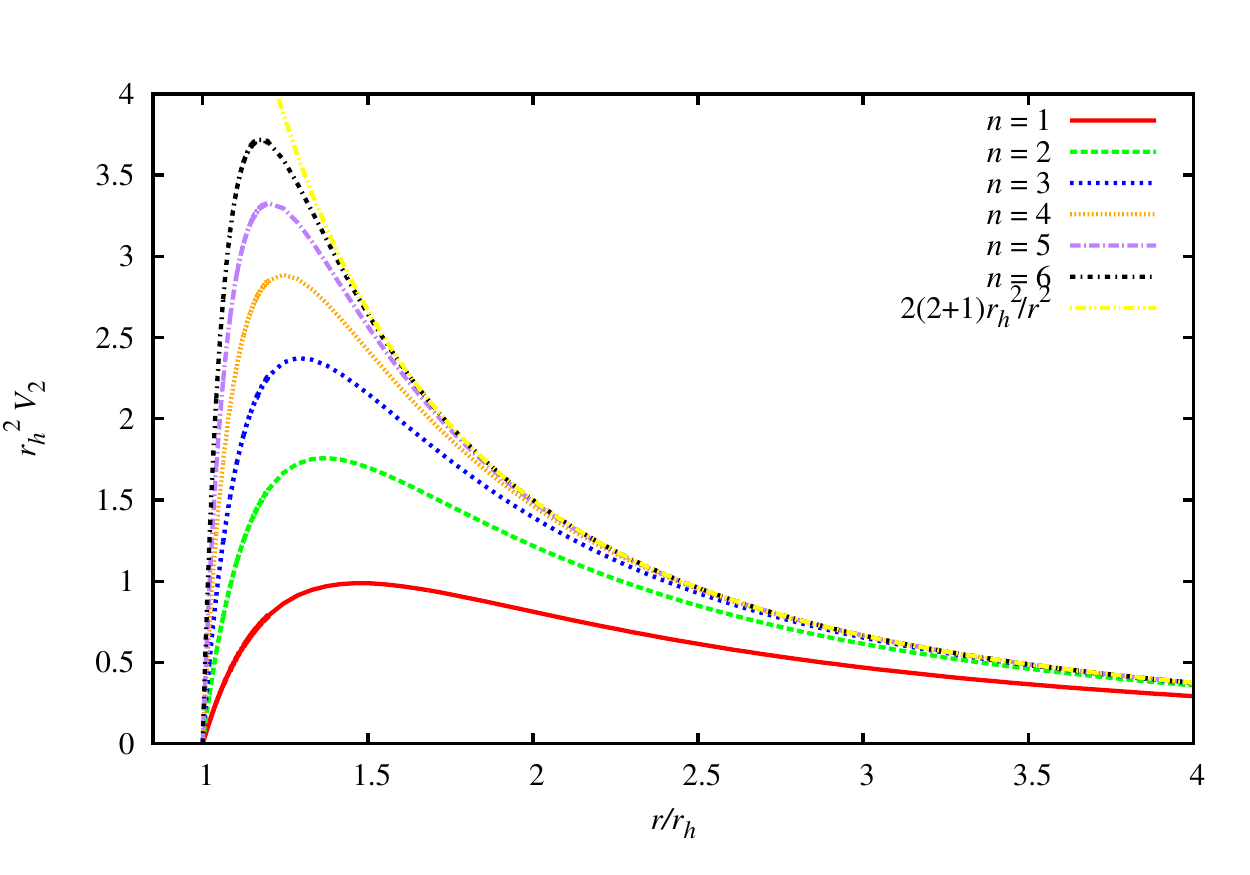}}
	\caption{Effective potential for (a) $l=0$ and (b) $l=2$ considering different values of $n$. Near the event horizon and at infinity, the effective potential goes to zero. In (b) we also plot the non-relativistic centrifugal barrier, $l(l+1)/r^2$, and we note that as $r$ increases, the effective potential tends to it.}
	
	\label{potencial-l02}
\end{figure}

Figure~\ref{potencial-l02} shows the effective potential for $l=0$ and $l=2$ and different values of $n$. It goes to zero near the event horizon and at infinity.

In order to investigate the scattering process we assume purely ingoing waves near the event horizon, so that the radial solution in this region can be written as
\begin{equation}
\psi^{(n)}_{\omega l}(r_*)\approx A_{\omega l}^{(n)\mathrm{tr}}\textrm{e}^{-i\omega r_*},
\label{hor-sol}
\end{equation}
where $A_{\omega l}^{(n)\mathrm{tr}}$ is a coefficient which modulus square determines the amount of wave absorbed by the black hole.

Far from the event horizon, we have $V_l\approx l(l+1)/r_{*}^2$, what produces
\begin{eqnarray}
\psi^{(n)}_{\omega l}(r_*) & \approx & \omega r_*\left[(-i)^{l+1}A_{\omega l}^{(n)\mathrm{in}}h_l^{(2)}(\omega r_*) \right. \nonumber \\ 
&&\left. + i^{l+1}A_{\omega l}^{(n)\mathrm{ref}}h_l^{(1)}(\omega r_*)\right].
\label{inf-hankel-sol}
\end{eqnarray}
In Eq.~\eqref{inf-hankel-sol}, $h_l^{(1)/(2)}(\cdot)$ are the spherical Hankel functions \cite{abramowitz} and the coefficients $A_{\omega l}^{(n)\mathrm{in/ref}}$ are related to the amount of incident (in) and reflected (ref) waves.

The phase shifts $\delta_l^{(n)}$ are defined by~\cite{gottfried}
\begin{equation}
\delta_l^{(n)}(\omega)=\frac{1}{2i}\ln\left[(-1)^{l+1}\frac{A_{\omega l}^{(n)\mathrm{ref}}}{A_{\omega l}^{(n)\mathrm{in}}}\right],\label{phaseshift.def}
\end{equation}
and the scattering amplitude by 
\begin{equation}
f^{(n)}_\omega(\theta)=\frac{1}{2i\omega}\sum_{l=0}^{\infty}(2l+1)\left(\textrm{e}^{2i\delta_l^{(n)}}-1\right)P_l(\cos\theta).
\label{fdeomega}
\end{equation}

Here we consider a monochromatic plane wave impinging upon the black hole. Such wave can be described as a composition of partial waves given by Eq.~\eqref{eq:decomposition}. In this case, the differential scattering cross section is directly obtained from $f^{(n)}_\omega(\theta)$ by
\begin{equation}
\frac{d\sigma^{(n)}}{d\Omega}=\big|f^{(n)}_\omega(\theta)\big|^2.
\end{equation}

The definitions of the scattering $\sigma^{(n)}_{\mathrm{el}}$, absorption $\sigma^{(n)}_{\mathrm{abs}}$ and total cross sections $\sigma^{(n)}_{\mathrm{tot}}=\sigma^{(n)}_{\mathrm{el}}+\sigma^{(n)}_{\mathrm{abs}}$ in terms of the phase shifts are:
\begin{equation}
\sigma^{(n)}_{\mathrm{el}}=\sum_{l=0}^{\infty}\sigma^{(n)}_{l,\mathrm{el}}=\frac{\pi}{\omega^2}\sum_{l=0}^{\infty}(2l+1)\left|\textrm{e}^{2i\delta_l^{(n)}}-1\right|^2; \label{scat.sec.}
\end{equation}
\begin{equation}
\sigma^{(n)}_{\mathrm{abs}}=\sum_{l=0}^{\infty}\sigma^{(n)}_{l,\mathrm{abs}}=\frac{\pi}{\omega^2}\sum_{l=0}^{\infty}(2l+1)\left[1-\left|\textrm{e}^{2i\delta_l^{(n)}}\right|^2\right];\label{abs.sec.}
\end{equation}
\begin{equation}
\sigma^{(n)}_{\mathrm{tot}}=\frac{2\pi}{\omega^2}\sum_{l=0}^{\infty}(2l+1)\left[1-\mathrm{Re}\left(\textrm{e}^{2i\delta_l^{(n)}}\right)\right],\label{tot.sec.}
\end{equation}
where $\sigma^{(n)}_{l,\mathrm{el}}$ and $\sigma^{(n)}_{l,\mathrm{abs}}$ are the respective partial scattering and absorption cross sections.

In this paper we determine the phase shifts via Born approximation, that is valid for $l\gg l_c\sim \omega b_c$ and via a numerical approach, valid for arbitrary values of the frequency. 

The Born approximation method can be used to compute phase shifts when the scattered wave is weakly modified by the target \cite{gottfried}. Therefore, results using this method are valid on the weak-field regime and for the cases $n > 1$. In order to compute the phase shifts via Born approximation, let us make the substitution $\psi_{\omega l}(r)=f^{-1/2}X(r)$ in Eq.~\eqref{r.radial-eq}, so that we obtain:
\begin{equation}
\frac{d^2}{dr^2}X+\frac{1}{f^2}\left[\omega^2-V_l(r)+\frac{f'(r)^2}{4}-\frac{f(r)f''(r)}{2}\right]X=0,
\end{equation}
where the prime denotes differentiation with respect to $r$.
Expanding the equation above in powers of $1/r$, we obtain:
\begin{equation}
\frac{d^2}{dr^2}X+S(r)X=0,
\end{equation}
where
\begin{equation}
S(r)=\sum_{k=0}^{\infty}\left[\omega^2(k+1)+\frac{\alpha_k^{(n)}-l(l+1)}{r^2}\right]\left(\frac{r_h}{r}\right)^{nk},\label{aproxim}
\end{equation}
with $\alpha_k^{(n)}$ being constants, in particular, $\alpha_0^{(n)}=0$. In Table~\ref{tb:alphak} we present the values of $\alpha_k^{(n)}$ for the other first three terms of the series for $n = 1\ldots 6$.
\begin{table}[htb!]
	\centering
	\caption{Values of $\alpha_k^{(n)}$ for $k=1,2$ and $3$ and $n = 1\ldots 6$.}\label{tb:alphak}
 \begin{tabular}{cccc}
 	\hline\hline $n$ & $\alpha_1^{(n)}$ & $\alpha_2^{(n)}$ & $\alpha_3^{(n)}$ \\ 
 	\hline 1 & 0 & 1/4 & 1/2 \\ 
 	2 & 1 & 2 & 3 \\ 
 	3 & 3 & 21/4 & 15/2 \\ 
 	4 & 6 & 10 & 14 \\ 
 	5 & 10 & 65/4 & 45/2 \\ 
 	6 & 15 & 24 & 33 \\ 
 	\hline\hline 
 \end{tabular} 
\end{table}

The case $n=1$ is asymptotically similar to the Coulomb potential and therefore cannot be treated via Born approximation~\cite{Futterman_etal-1988,JMP_1976-Norma}. For $n=2$, we define
\begin{equation}
U_2(r)=-\left[2\omega^2+\frac{3\omega^2r_h^2+\alpha_1^{(2)}-l(l+1)}{r^2}\right]\frac{r_h^2}{r^{2}}+\mathcal{O}\left(r^{-6}\right),\label{Uder2}
\end{equation}
while for $n\geqslant2$ we define
\begin{equation}
U_{n}(r)=-\left[2\omega^2+\frac{\alpha_1^{(n)}-l(l+1)}{r^2}\right]\frac{r_h^n}{r^{n}}+\mathcal{O}\left(r^{-2n}\right),\label{Uder}
\end{equation}
so that the radial equation~\eqref{aproxim}, for $n\geq2$, can be written as
\begin{equation}
\frac{d^2}{dr^2}X+\left[\omega^2-\frac{l(l+1)}{r^2}-U_n(r)\right]X=0.
\end{equation}

By applying the Born approximation formula \cite{gottfried}, namely
\begin{equation}
\delta_l^{(n)}\approx -\omega \int_{0}^{\infty} r^2[j_l(\omega r)]^2U_{n}(r)dr,\label{born-approx}
\end{equation}
where $j_l(\cdot)$ are the spherical Bessel functions of first kind \cite{abramowitz}, and the identity
\begin{equation}
\int_{0}^{\infty}x^{-\nu}[j_l(x)]^2dx=\frac{\sqrt{\pi}}{4}\frac{\Gamma\left(\frac{1+\nu}{2}\right)}{\Gamma\left(1+\frac{\nu}{2}\right)}\frac{\Gamma\left(l+\frac{1-\nu}{2}\right)}{\Gamma\left(l+\frac{3+\nu}{2}\right)}, \label{ident}
\end{equation}
valid for $l>(\nu-1)/2$, and $\nu\geqslant 0$, with $\Gamma(\cdot)$ being the gamma function, we find the phase shifts $\delta_l^{(n)}$. Expanding the result in terms of $l+1/2$, and keeping only the leading term, we obtain:
\begin{equation}
\delta_l^{(n)}\approx\frac{\sqrt{\pi}}{2(n-1)}\frac{\Gamma\left(\frac{n+3}{2}\right)}{\Gamma\left(\frac{n+2}{2}\right)}\frac{(\omega r_h)^{n}}{(l+1/2)^{n-1}}.
\label{delta-l}
\end{equation}

Through the Ford and Wheeler's semi-classical description of scattering \cite{Ann.Phys.Ford.Wheeler} we can make a connection between the phase shifts~\eqref{delta-l} and the weak-field deflection~\eqref{wf_def}. By doing so, we can verify that the results are consistent with each other.

\section{Numerical analysis}\label{Sec:Num.Analysis}

In order to determine the cross sections for arbitrary values of the scattering angle and wave frequency, we have to find numerically the coefficients $A_{\omega l}^{(n)\mathrm{in}}, A_{\omega l}^{(n)\mathrm{ref}}$ and $A_{\omega l}^{(n)\mathrm{tr}}$ from Eqs.~\eqref{hor-sol} and \eqref{inf-hankel-sol} by solving the radial equation \eqref{r.radial-eq}. We start close to the event horizon $r=(1+\epsilon)r_h$, where $\epsilon$ is usually chosen to be $10^{-4}$, with the boundary condition \eqref{hor-sol}, which can be improved with a series expansion like:
\begin{equation}
\psi_{\omega l}^{(n)}(r)=\mathrm{e}^{-i\omega r_*}\sum_{q=0}^{q_m}a_q(r-r_h)^q,\label{hor-approx}
\end{equation}
where $a_0 = A_{\omega l}^{(n)\mathrm{tr}}$ and higher values of $q_m$ imply more precise results.

We have checked that computations for $q_m=2$ and higher do not result in appreciable numerical differences for values of $\psi_{\omega l}^{(n)}$ taken far from the event horizon (asymptotic plane region), typically chosen as $r=r_m\sim(400+5l)r_h$, for $n\geq3$. For $n=1$ and $n=2$ a larger value of $r_m$ is needed to reach the asymptotic plane region. We input $l$-dependent values for $r_m$ because as the larger the value of $l$ is, further from the black hole the related partial wave passes as it possesses a higher impact parameter. Finally, obtaining the phase shifts demands matching the numerical values of $\psi_{\omega l}^{(n)}$ and $d\psi_{\omega l}^{(n)}/dr$ with their asymptotic solutions, Eq.~\eqref{inf-hankel-sol}. 

\begin{figure}[htb!]
\subfigure[\ $n = 1, 2, 3$.]{\includegraphics[width=\columnwidth]{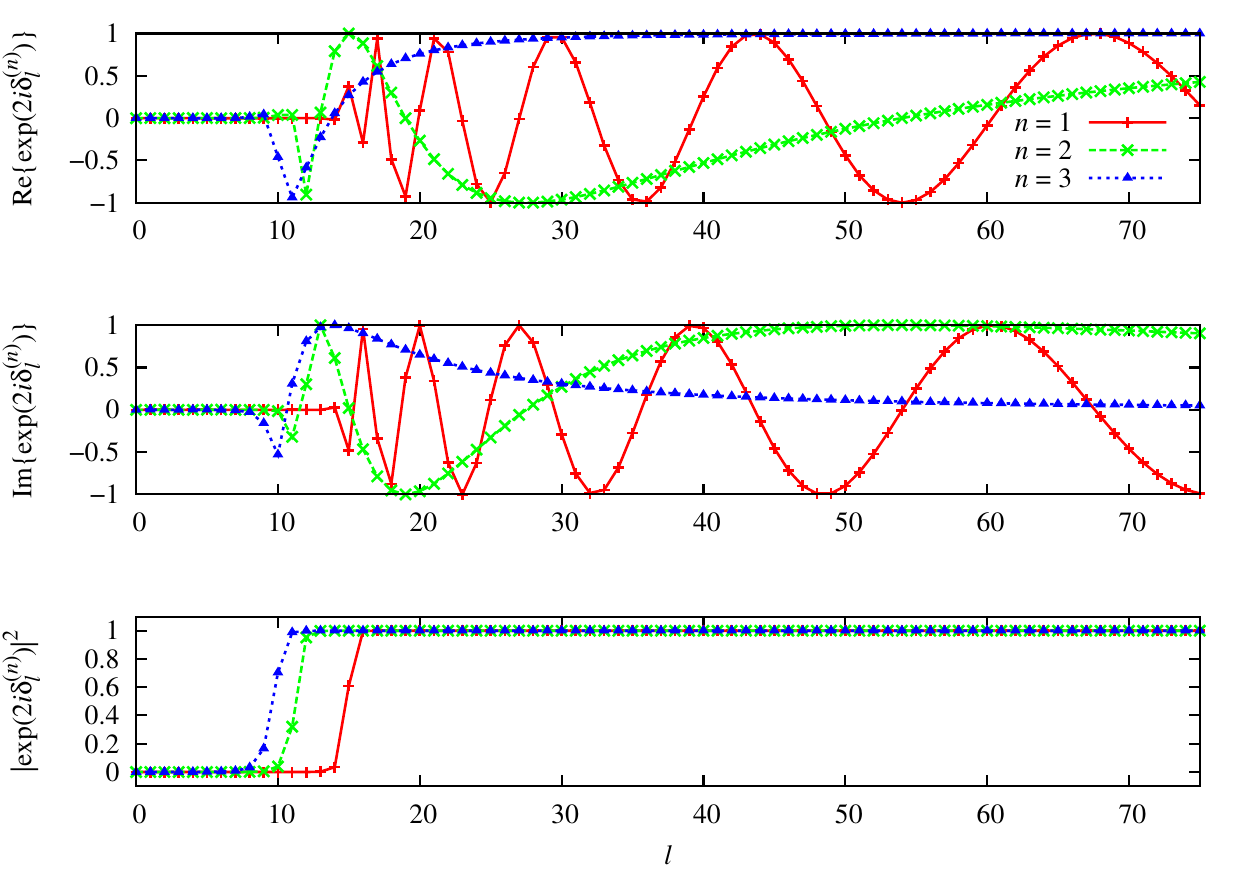}}
\subfigure[\ $n = 4, 5, 6$.]{\includegraphics[width=\columnwidth]{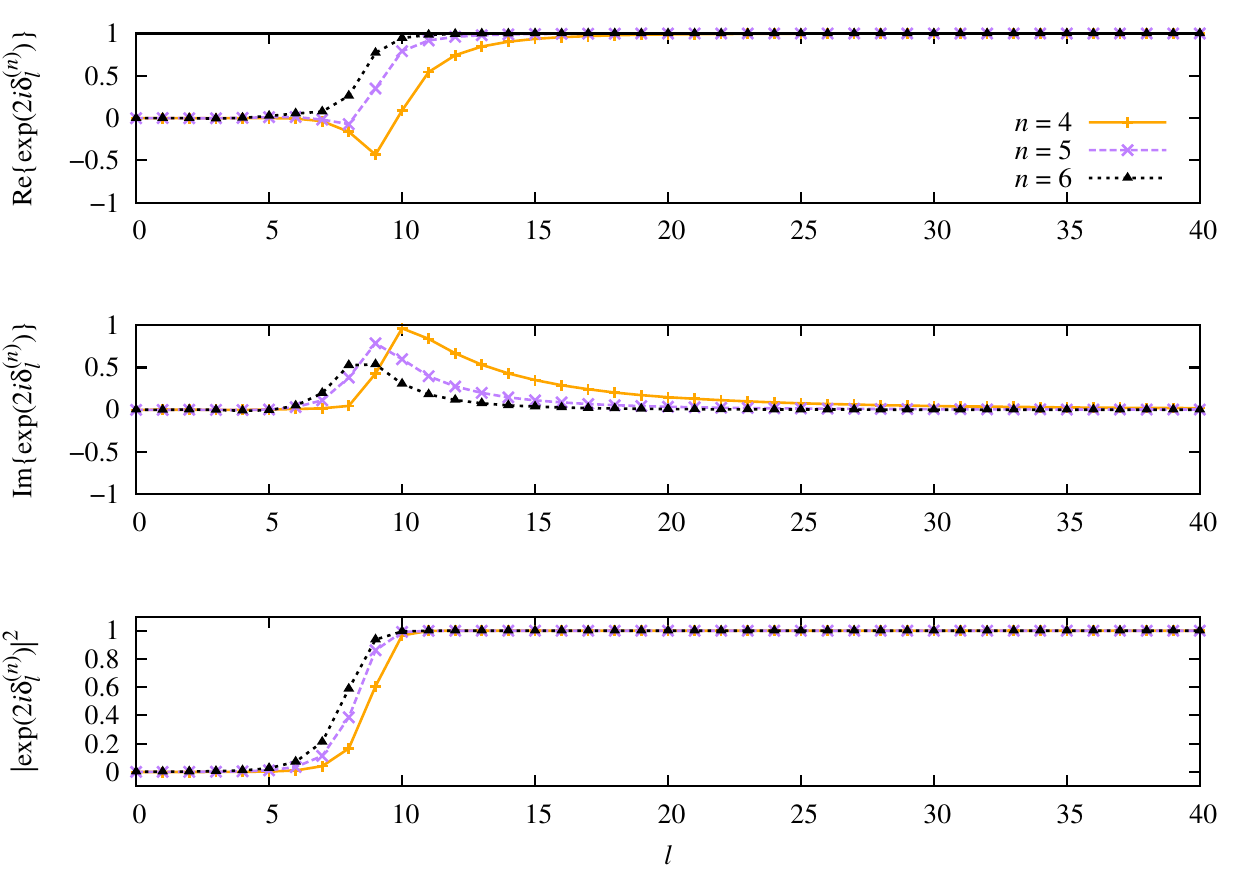}}
\caption{Real and imaginary parts of $\mathrm{e}^{2i\delta_m^{(n)}}$ as well as its quadratic module in the case $\omega r_h=6.0$ (a) for $n=1,2, 3$; and (b) for $n=4, 5$ and $6$.}\label{fig:phaseshift}
\end{figure}

Figure~\ref{fig:phaseshift} shows the quadratic module as well as the real and imaginary parts of $\mathrm{e}^{2i\delta_m^{(n)}}$ for some values of $n$ in the case $\omega r_h=6.0$. For $l\ll\omega b_c$, most of the partial waves are absorbed. In the large-$l$ limit, the phase shifts are real (in accordance with Eq.~\eqref{delta-l}), so that most of the partial waves are scattered. For intermediate values of $l$, waves are partially absorbed and partially reflected ($0 <  \left| \mathrm{e}^{2i\delta_m^{(n)}} \right|^2 < 1$).

Once the numerical phase shifts are computed, we can obtain the scattering amplitudes. For $n=1$, it is known that a convergence method is needed to compute the scattering cross section, since the summation given by Eq.~\eqref{fdeomega} converges poorly for $\theta > 0$~\cite{PhysRev.95.500}. The same is true for $n=2$ and $n=3$. Particularly, in such cases, the scattering amplitude diverges in the forward direction, where the differential scattering cross section is infinite. Additionally, in the case for two extra dimensions ($n = 3$) this divergence comes from the real part of the scattering amplitude, while the imaginary part converges (which can be shown using the second Born approximation \cite{gottfried}). However, for $n\geq4$, the scattering amplitude is convergent in all directions. In order to understand this, let us separate the scattering amplitude as
\begin{equation}
 f^{(n)}_\omega(\theta) \approx f^{(n)}_{\omega,\mathrm{num}}(\theta)+\frac{1}{\omega}\sum_{l=l_{M}}^{\infty}(2l+1)\delta_l^{(n)}P_l(\cos\theta),
 \label{sep_amp}
\end{equation}
where $f^{(n)}_{\omega,\mathrm{num}}(\theta)$ is obtained numerically, $l_M$ is chosen in such way that the phase shifts obtained numerically and via Born approximation are very close to each other, and the phase shifts in the second term of the right hand side of Eq.~\eqref{sep_amp} are given by Eq.~\eqref{delta-l}. This series is absolutely convergent for all $\theta$ only if $n\geq4$, once $|P_l(\cos\theta)| \le 1$ and the summation coefficients fall off with $(l+1/2)^{2-n}$. Yet, once $P_l(1) = 1$, the scattering amplitude diverges for $\theta = 0$ in the cases $n\le 3$.

For the case $n=3$ an interesting fact is observed; as stated in the previous paragraph, the differential scattering cross section diverges in the limit $\theta\to0$. Nevertheless, the summations for both scattering and total cross sections [Eqs.~\eqref{scat.sec.} and \eqref{tot.sec.}] are indeed finite. We can infer this from Eq.~\eqref{scat.sec.} with the phase shift calculated via Born approximation:
\begin{equation}
\sigma^{(n)}_{\mathrm{el}} \approx \sigma^{(n)}_{\mathrm{el,num}}+\frac{2\pi}{\omega^2}\sum_{l=l_{M}}^{\infty}(2l+1)\left[\delta_l^{(n)}\right]^2,
\label{sep_f}
\end{equation}
which make us conclude that the sum converges for $n>2$. The same analysis can be done for $\sigma^{(n)}_{\mathrm{tot}}$.

\section{Results}\label{Sec:Results}

In this section we present our numerical results as well as their comparisons with the analytical approximations. In order to compute the differential scattering cross section numerically, we apply a convergence method to obtain the scattering amplitude expanded in terms of partial waves, Eq.~\eqref{fdeomega}, for the cases $n = 1, 2, 3$. For the cases $n \ge 4$, we use Eq.~\eqref{sep_amp} instead, with $l_M$ within the regime of validity of the Born approximation.

\begin{figure}[htb!]
		\centering
		\includegraphics[width=\columnwidth]{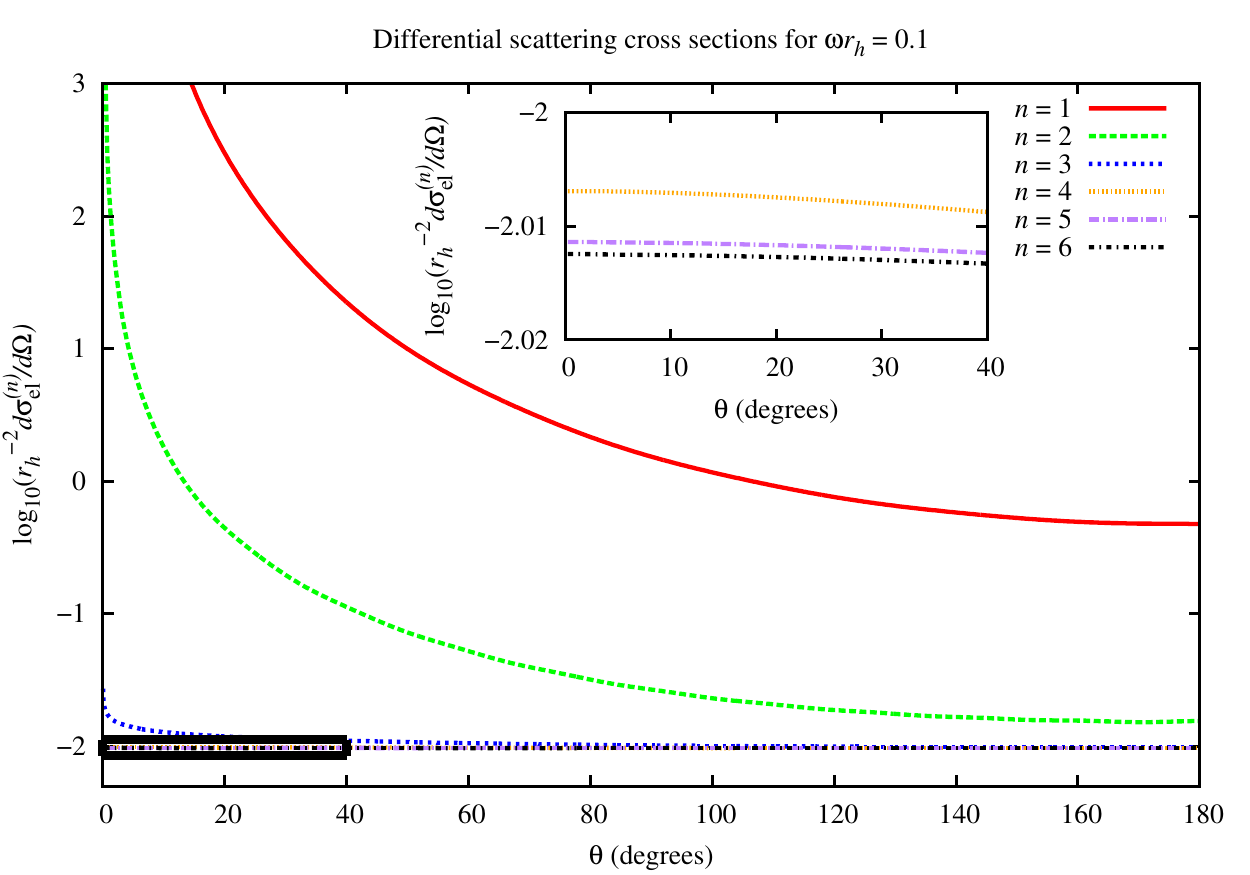}
		\caption{Differential scattering cross sections for the massless scalar field on the Schwarzschild brane for $\omega r_h=0.1$. For $n<4$ we have a divergence in $\theta=0$. We see that the amount of scattered flux falls rapidly with the increase of $n$. The zoom-in area shows the cases $n = 4,5,6$ in the near-forward directions.
		}\label{fig:scatt001}
\end{figure}

In Fig.~\ref{fig:scatt001} we show the differential scattering cross sections for $\omega r_h=0.1$ and $n = 1 \ldots 6$. As $n$ increases, the scattering amplitude decreases. This is a expected result, as the higher is the dimensionality of the spacetime, the faster $f(r)$ tends to 1 far from the black hole. As we can clearly see in the zoom-in box at the figure, for the cases in which $n>3$ we have finite differential scattering cross section in the forward direction.

\begin{figure}[htb!]
	\centering
	\subfigure[]{\label{fig:scatt6a}\includegraphics[width=\columnwidth]{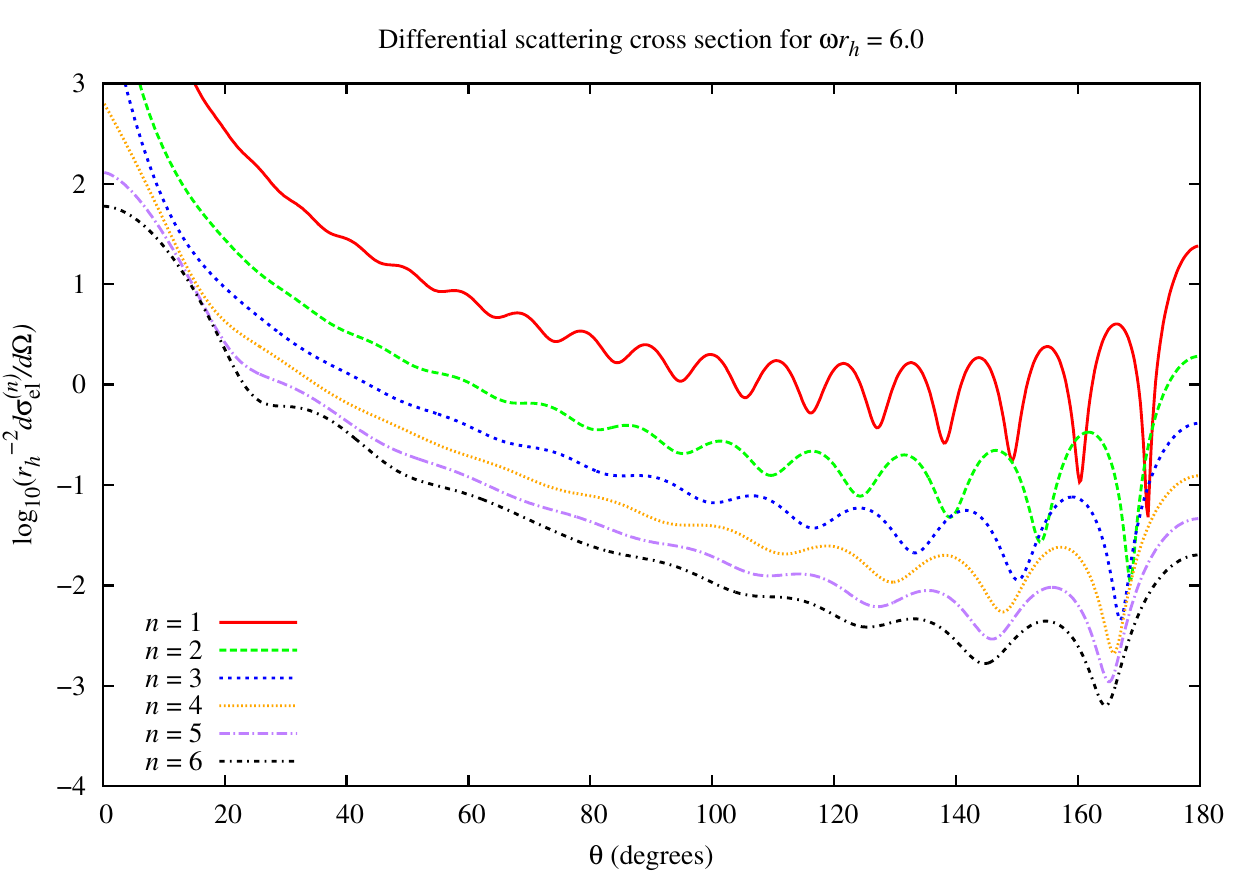}}
	\subfigure[]{\label{fig:scatt10b}\includegraphics[width=\columnwidth]{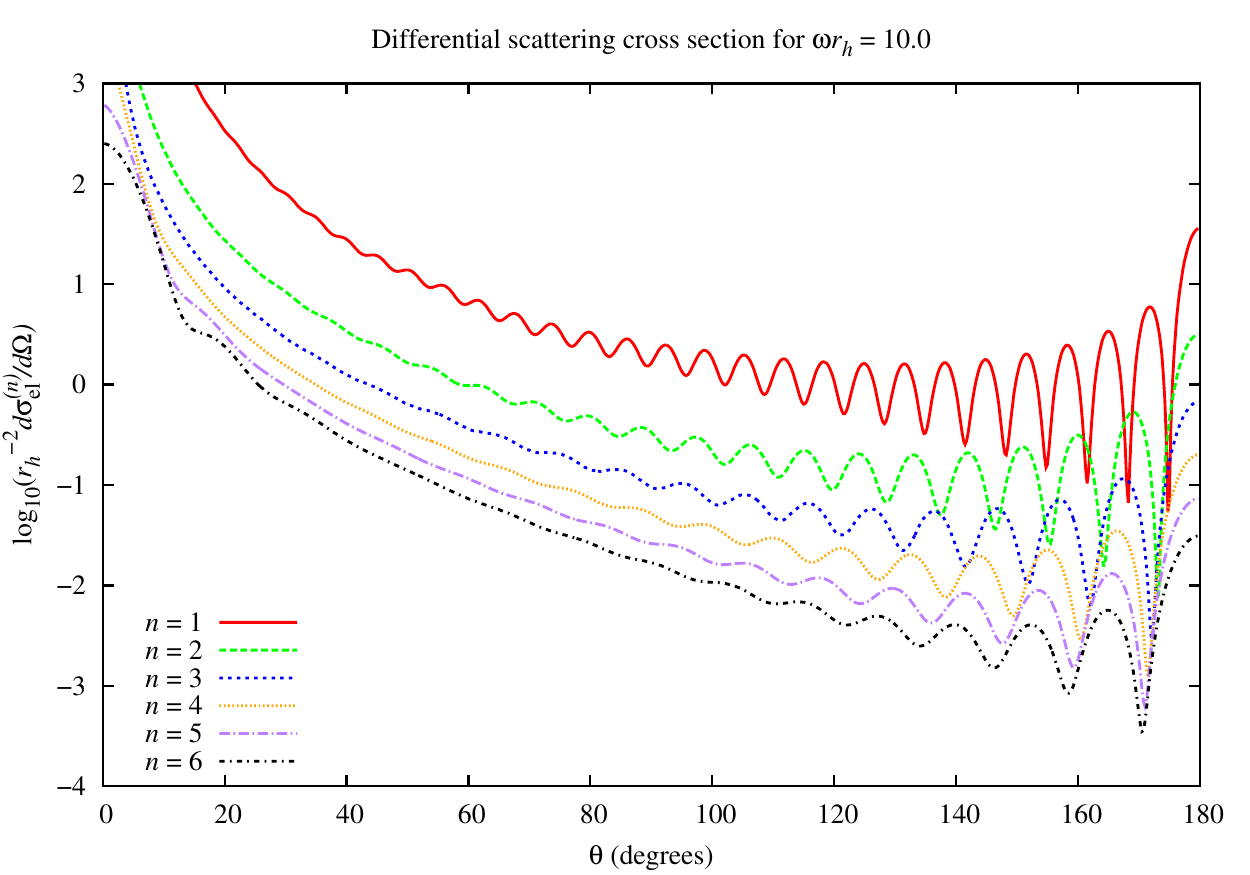}}
	\caption{Differential scattering cross section for (a) $\omega r_h=6.0$ and (b) $\omega r_h=10.0$. For both [(a) and (b)] frequency values, the differential scattering cross section becomes smaller as $n$ increases. The number of local maxima (fringes of interference) increases with the increase of the frequency or the decrease of the dimensionality of the spacetime.}
	\label{fig:scatt6e10}	
\end{figure}

\begin{figure}[htb!]
	\centering
	\includegraphics[width=\columnwidth]{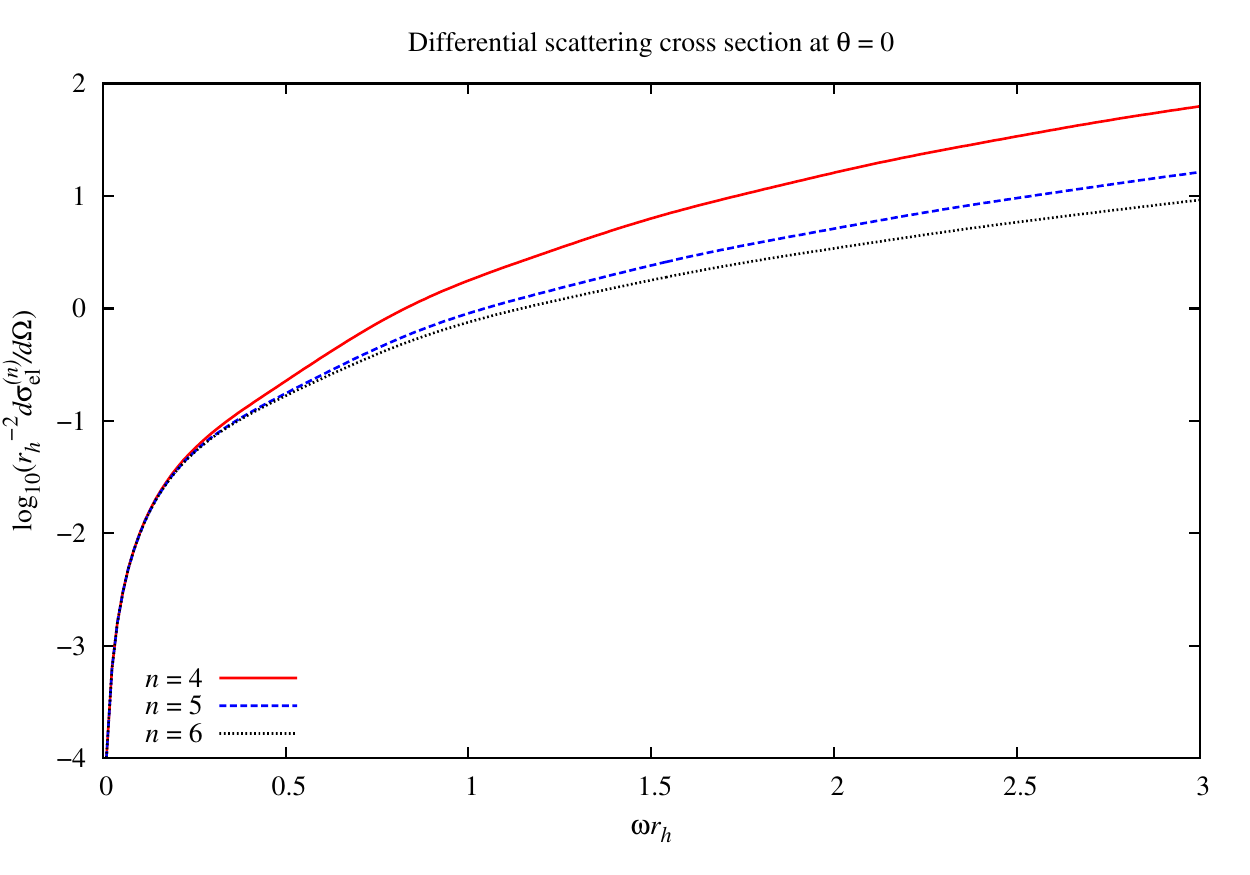}
	\caption{Differential scattering cross sections in the forward direction ($\theta=0$) for $n=4, 5$ and $6$ as function of $\omega r_h$. These cross sections are finite at $\theta=0$, and as $\omega r_h$ increases, they increase rapidly.}\label{fig:dscs-theta0}
\end{figure}

Figure~\ref{fig:scatt6a} presents the differential scattering cross sections for $\omega r_h=6.0$, where we observe the expected divergence of the scattering cross section in the forward direction for $n<4$, while it remains finite for $n\geq4$. The differential scattering cross sections for $\omega r_h=10.0$ are presented in Fig.~\ref{fig:scatt10b}. By comparing Figs.~\ref{fig:scatt6a} and~\ref{fig:scatt10b}, we see that the overall scattered flux increases in the forward direction for $n>3$ as we increase the wave frequency. Therefore, considering Eq.~\eqref{scatter-high-frenq}, we conclude that there must be infinity scattered flux in the forward direction for $n \ge 4 $ only in the limit $\omega r_h \to \infty$. This can be checked directly in Fig.~\ref{fig:dscs-theta0}, where we see the differential scattering cross sections at $\theta=0$ for $n\geq4$.

By comparing Figs.~\ref{fig:scatt001} and \ref{fig:scatt6e10}, we can infer that the number of local maxima for the differential scattering cross section (fringes) increases with the increase of the frequency or the decrease of the dimensionality of the spacetime.

\begin{figure}[htb!]
	\centering
	\includegraphics[width=\columnwidth]{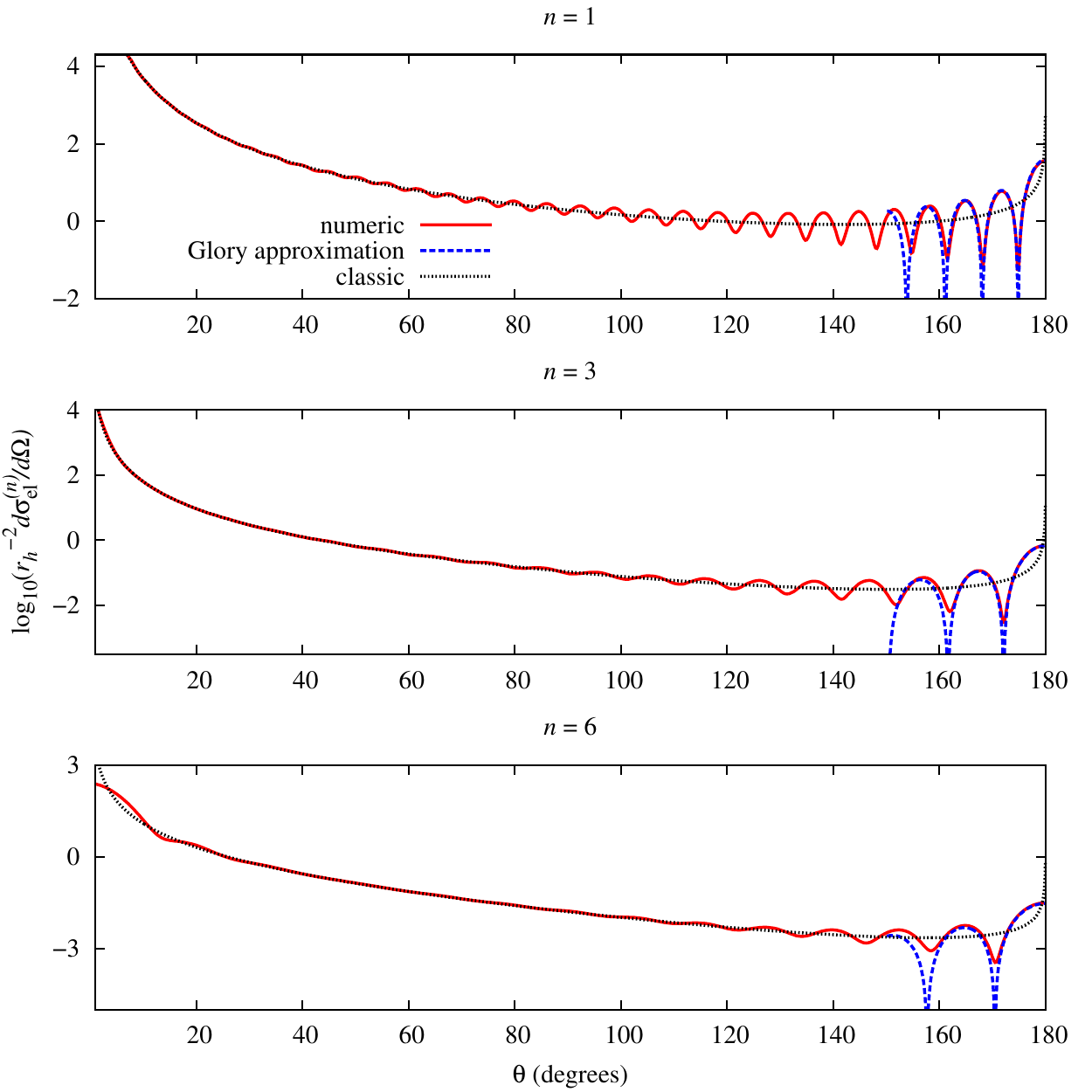}
	\caption{Comparison of the differential scattering cross sections with their classical limits and the analytic glory approximation with $\omega r_h=10.0$, for $n=1,3$, and $6$. For any value of the dimension parameter $n$, the glory rings form a good approximation near $\theta=\pi$, while the geodesic description for the scattering is a good approximation for small angles in the cases $n\leq 3$, but not $n>3$.}\label{fig:compar}	
\end{figure}

In Fig.~\ref{fig:compar}, we show for $n=1$, $n=3$ and $n=6$ the differential scattering cross sections obtained via geodesic analysis, Eq.~\eqref{eq.class.scat.}, via the glory approximation (with $\omega r_h=10.0$), Eq.~\eqref{eq.glory.scat.}, and numerically (with $\omega r_h=10.0$). As we can see, the glory approximation agrees very well with the numerical results for $\theta\approx\pi$, as we should expect. Once the differential scattering cross section is finite for $n\geq4$ as $\theta \to 0$, the classical limit is not a good approximation for small scattering angles in such cases.

\begin{figure}[htb]
	\centering
	\includegraphics[width=\columnwidth]{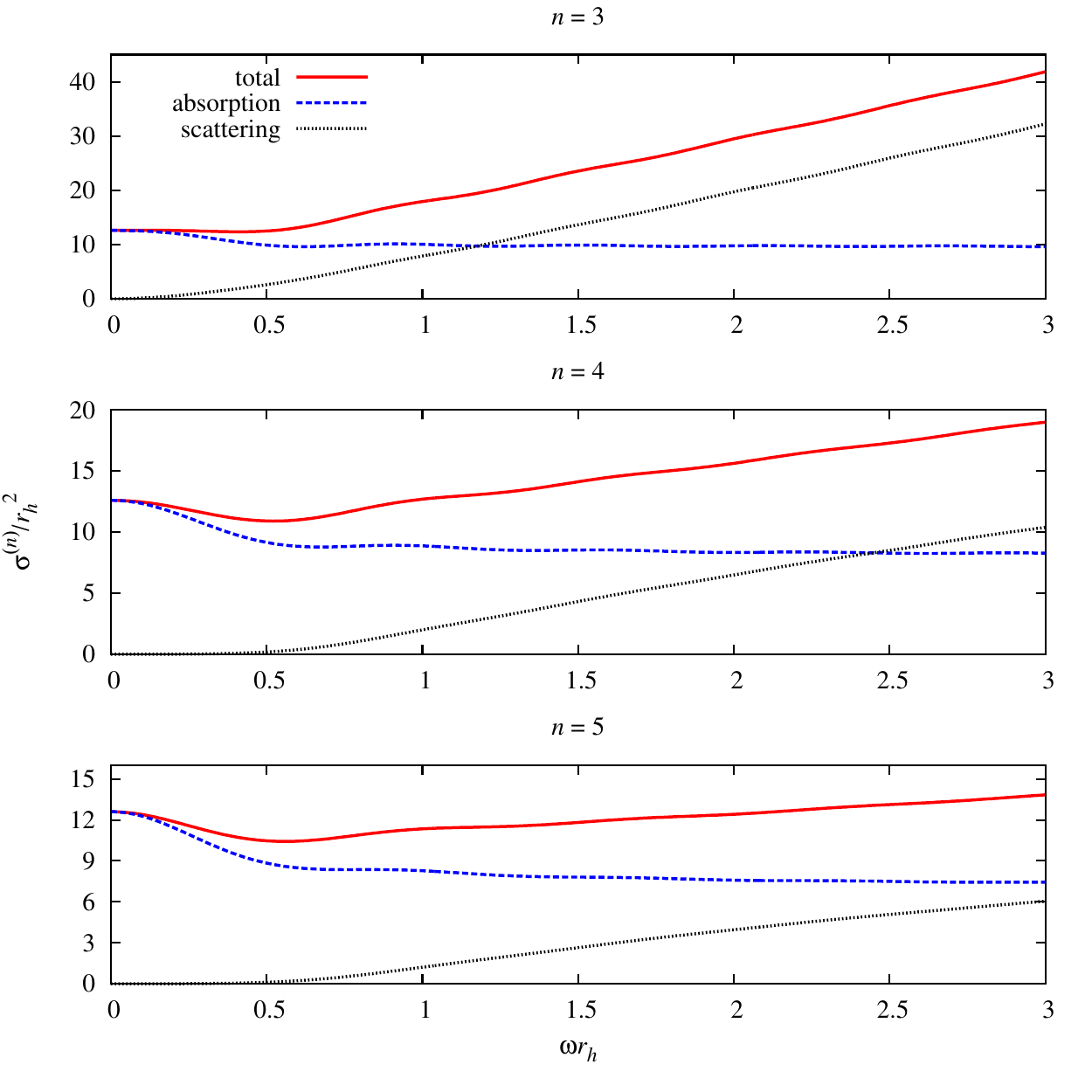}
	\caption{Scattering, absorption, and total cross sections for the cases $n = 3, 4$ and $5$ in frequency range $0<\omega r_h<3$.}\label{fig:cross.sections}
\end{figure}

In Fig.~\ref{fig:cross.sections} we show the absorption, scattering, and total cross sections for $n=3, 4$ and $5$ within the frequency interval $0<\omega r_h<3.0$. As highlighted in previous section, although the differential scattering cross section for $n=3$ diverges at $\theta=0$, we verify that the scattering cross section is finite.

For the values of $n$ above, $\sigma^{(n)}_{\mathrm{el}}$ increases with $\omega r_h$ in almost a linear fashion, noting that as $n$ increases its inclination decreases. The absorption cross sections $\sigma^{(n)}_{\mathrm{abs}}$ for low frequencies tend to the event horizon area (on the brane) $\sigma^{(n)}_{\mathrm{abs}}\approx4\pi r_h^2$ agreeing with the results for the low-frequency regime \cite{PhysRevD.66.024023,das1997,Jung:2005mr,atsushi-classquant,*higuchi2002cqg19_599} and tend to the geometric-optics value $\sigma^{(n)}_{\mathrm{abs}}\approx \pi b_c^2$ in the high-frequency limit. Like the scattering cross section, the absorption cross section gets smaller for higher values of $n$.

\section{Conclusion}\label{Sec:Conclusion}
We have investigated the scattering of the massless scalar field restricted to the 3-brane of a $(n+3)$-dimensional Schwarzschild black hole. Numerical results were presented for the cases $n = 1,\ldots, 6$, but our analysis could be applied to obtain results for any value of $n$.

In order to compute the differential scattering cross section, we used the partial-wave method. We applied a convergence method to compute the scattering amplitude in terms of partial waves for the cases $n = 1,2,3$. For $n = 4,5,6$, the sum of the scattering amplitude is convergent for all scattering angles. This was shown using the phases shifts computed via the Born approximation, Eq.~\eqref{delta-l}. In such cases, the analytical phase shifts were used together with the numerical ones to generate precise results.

Our results show that there are branes with two different behaviors when considering the scattering in the forward direction. For $n = 1,2,3$, we obtained infinite flux scattered in the $\theta \to 0$ limit, while the flux is finite for $n \ge 4 $. It is worth mentioning that the 3-brane metric of a 7-dimensional Schwarzschild black hole ($n = 4$) coincides with the metric for the canonical acoustic hole, whose scattering properties have been studied in Refs.~\cite{crispino2007prd76_107502,PhysRevD.79.064014}.

In the case $n = 3$, the differential scattering cross section is divergent in the forward direction, while the total cross section is finite. At first glance, it may seem that this result is inconsistent with the optical theorem, which can be expressed as $\sigma_\text{tot}^{(n)} = (4\pi/\omega) \text{Im} f_\omega^{(n)} (0)$~\cite{gottfried}. We showed analytically in Sec.~\ref{Sec:Num.Analysis}, using the Born approximation, that the scattering amplitude diverges in the case $n = 3$ for $\theta = 0$ [cf. Eq.~\eqref{sep_amp}]. However, the first order in the Born approximation gives only a real contribution to the scattering amplitude in the present case. To resolve the issue, one should consider the second Born approximation~\cite{gottfried}. We have verified numerically that the imaginary part of the scattering amplitude at $\theta = 0$ for $n = 3$ is indeed finite and thus compatible with the results presented in Fig.~\ref{fig:cross.sections}, which were obtained via Eq.~\eqref{tot.sec.}.

We have compared the numerical results with analytical results obtained via geodesic analysis and via glory approximation. The glory approximation, Eq.~\eqref{eq.glory.scat.}, agreed very well with all numerical results for $\theta \lesssim 180^\circ$. The classical scattering cross section agreed very well with the numerical results for the cases $n= 1, 2, 3$ even for intermediate values of frequency, but not with the cases $n \ge 4$, for which finite differential scattering cross sections in the forward direction were observed.

Finally, we should mention that the results presented here are only valid for black holes that are much smaller than the extra dimensions. For such black holes, a more realistic treatment should include the dynamics of the scattering process, as well as the variation of mass due to the Hawking radiation. It would also be interesting to study the cases in which the black holes have sizes comparable to the extra dimensions. This problem is currently under investigation.

\acknowledgments

The authors would like to thank Lu\'{\i}s C. B. Crispino, George E. A. Matsas and Sam R. Dolan for useful discussions and suggestions and to Conselho Nacional de Desenvolvimento Cient\'ifico e Tecnol\'ogico (CNPq) and Coordena\c{c}\~ao de Aperfei\c{c}oamento de Pessoal de N\'ivel Superior (CAPES) for partial financial support. C. I. S. M. dedicates this work to the memory of Silvia C. Santa Rosa.

\end{document}